\documentstyle[aps,pre,graphicx,multicol,amssymb,amsmath]{revtex}

\draft
\tighten

\newcommand{\ie}{{\it i.e.\/}}
\newcommand{\Lp}{{\cal L}_p} %persistent length scale
\newcommand{\LD}{{\cal L}_D} %diffusive  length scale
\newcommand{\Lc}{{\cal L}_c} %characteristic length scale

\begin{document}

\title{Persistence in Cluster--Cluster Aggregation
}
\vspace{0.8cm}
\author {E.~K.~O.~Hell\'{e}n and M.~J.~Alava} 
\address{Helsinki University of Technology, Laboratory of
Physics, P.O.Box 1100, FIN-02015 HUT, Finland}

\date{\today}

\maketitle

\begin{abstract}
Persistence is considered in diffusion--limited cluster--cluster aggregation,
in one dimension and when the
diffusion coefficient of a cluster depends on its size $s$ as $D(s)
\sim s^\gamma$. 
The empty and filled site persistences are defined as
the probabilities, that a site
has been either empty or covered by a cluster all the time whereas the
cluster persistence gives the probability of a cluster to remain intact.
The filled site one is nonuniversal. The empty site and
cluster persistences are found to be universal, as
supported by analytical arguments and simulations. 
The empty site case decays algebraically with the
exponent $\theta_E = 2/(2 - \gamma)$. 
The cluster persistence is related to the small $s$ behavior
of the cluster size distribution and behaves also algebraically for
$0 \le \gamma < 2$ while for $\gamma < 0$ the behavior is stretched
exponential. In the scaling limit $t \to \infty$ and $K(t) \to \infty$
with $t/K(t)$  
fixed the distribution of intervals of size $k$ between persistent regions
scales as $n(k;t) = K^{-2} f(k/K)$, where $K(t) \sim t^\theta$ is the
average interval size and $f(y) = e^{-y}$. 
For finite $t$ the scaling is poor for $k \ll t^z$, due to the
insufficient separation of the two length scales: the distances between
clusters, $t^z$, and that between persistent regions, $t^\theta$. 
For the size distribution of persistent regions the time and 
size dependences separate, the latter being independent of the
diffusion exponent $\gamma$ but depending on the initial
cluster size distribution. 
 
\end{abstract}


%05.40.-a Fluctuation phenomena, random processes, noise, and Brownian
%       motion 
%82.20.Mj Nonequilibrium kinetics 
%05.50.+q Lattice theory and statistics (Ising, Potts, etc.) (see also
%       64.60.Cn Order- disorder transformations and statistical
%       mechanics of model systems and 75.10.Hk Classical spin models)
%05.70.Ln Nonequilibrium and irreversible thermodynamics (see also
%       82.40.Bj Oscillations, chaos, and bifurcations in physical chemistry
%       and chemical physics) 
%02.50.Ey Stochastic processes

\begin{multicols}{2}[]

\section{Introduction} \label{intro}

Persistence in dynamical systems has attained considerable
interest recently~\cite{Majumdar:CurrSciIndia}. It was originally
introduced for a simple diffusion
process~\cite{Majumdar:PRL77,Derrida:PRL77} but since then it has been
studied, for example, in spin
systems~\cite{Bray:PRE62,Derrida:PRE54,Krapivsky:PRE56,Ben-Naim:JSP93,Sire:PRE52,Spirin:PRE60,Jain:JPHYSA33,Fisher:PRL80,ODonoghue:CM2},
reaction-diffusion
systems~\cite{ODonoghue:PRE64,Manoj:,Ben-Naim:comment,Manoj:reply,Manoj:JPA33L,Manoj:JPA33,ODonoghue:CM1},
voter model~\cite{Ben-Naim:PRE53,Howard:JPA31}
and for
interfaces~\cite{Majumdar:PRL86,Kallabis:EPL45,Kallabis:PRE56,Toroczkai:PRE60}.
It has also been measured
experimentally~\cite{Tam:PRL78,Tam:EPL51,MarCos-Martin:PHYSICAA214,Wong:PRL86,Yurke:PRE56},
and a few exact results
exist ~\cite{Derrida:PRL75,Derrida:JSP85,Burkhardt:JPA26,Baldassarri:PRE59}.

The persistence is usually defined as the probability 
$P(t)$ that at a fixed point in space a fluctuating nonequilibrium
field $\phi(x;t)$ does 
not change sign upto time $t$, {\it i.e.\/}, the probability that
$\mbox{sgn}[\phi(x;t)-\langle \phi(x;t) \rangle]$ remains
unchanged. In many systems the probability decays
algebraically, $P(t) \sim t^{-\theta}$, with a nontrivial persistence
exponent $\theta$. The significance of the phenomenon stems from the fact
that the  exponent is not, in general, related  
to the usual static or dynamic exponents. This this in turn implies
that not necessarily all of the properties 
of the system are characterized by a single length scale.

In this article we study various definitions
of persistence in an aggregation process, in the disguise
of the one-dimensional diffusion--limited 
cluster--cluster aggregation model
(DLCA)~\cite{Meakin:PhysicaScripta46}, with each cluster   
diffusing with a size dependent diffusion coefficient, $D(s) \sim
s^\gamma$ with $\gamma<2$.  
This model is in many respects suitable for studying persistence as it
possesses a rich dynamical behavior but at the same time it is simple
enough to often allow for an analytical treatment~\cite{Spouge:PRL60}. 
Moreover in DLCA one meets immediately the possibility of
defining several persistent quantities, each
of which describes a different aspect of the aggregation process. 

The persistence probabilities considered in this work are 
i)
the probability of a cluster to remain unaggregated (cluster
persistence), 
ii) 
the probability that a site has been empty (empty site
persistence) and 
iii) 
filled (filled site persistence) upto time $t$. 
Notice that all these are {\em local} quantities. It is
possible, in analogy to e.g. the
contact process~\cite{Munoz:PRE63}, to work starting from definitions
that involve 
a global quantity like the average cluster size. Thus one could
compare the ensemble average of the cluster size 
to that in any particular sample and investigate
the sign changes of their difference. Likewise one could attempt
the same, a comparison to the average cluster size,
for all individual clusters~\cite{Deloubriere:JPA33}.
These would be, in practice, much harder to study numerically
than i) to iii) above which also have the pleasant aspect of
being, possibly, experimentally relevant.
The comparisons between the three quantities points out
the underlying mechanisms, which make some persistent
quantities to be universal while the others are not. 

Recently Manoj and Ray  
claimed that the dynamic exponent
associated with the growth of the intervals between two consecutive
persistent sites in a diffusion--annihilation model $A + A \to
\emptyset$ shows nonuniversal behavior, \ie, the exponent
depends on the initial concentration of
particles~\cite{Manoj:}. This started a
discussion~\cite{Ben-Naim:comment,Manoj:reply}, concerning for example
whether it matters if only the empty sites are taken to be
persistent at $t=0$ instead of the entire lattice. The problem of
universality remained unsolved. Later on, the same authors argued that
the reason might be the competition between the two length scales, the
diffusive scale $\LD$ and the persistence length scale
$\Lp$~\cite{Manoj:JPA33L,Manoj:JPA33}. Very 
recently, Bray and O'Donoghue studied a similar problem in
the $q$-state Potts model~\cite{Bray:PRE62}. They found no evidence on the
concentration dependence and showed, that the characteristic length
scale $\Lc$ controlling the distribution of the intervals between
persistent sites, is given by the maximum of the two:
$\Lc = \mbox{max}\{\LD,\Lp\}$. Furthermore they argued that the poor
scaling for small values of $q$ is due to the insufficient
separation of these two length scales. These issues are easier to
resolve in the DLCA where the length scale separation is more
pronounced.

First we discuss the three persistence definitions with size
independent diffusion coefficients  ($\gamma  =0$)
in order to clarify the universality of these quantities. 
The filled site persistence turns out to be nonuniversal whereas the two
others decay similarly independent of the concentration and initial
conditions. This difference between empty and filled site
persistence 
is in clear contrast to all the previous persistence studies of,
for example, coarsening in the Ising model~\cite{Ben-Naim:JSP93},
where the symmetry between up and down spins is not broken. 

Thereafter we concentrate on the two universal 
persistent quantities and consider the influence of size dependent diffusion
($\gamma \neq 0$). We discover that it is possible to have
situations, in which the other persistence decays faster than any
power-law in time whereas the other shows algebraic behavior. 
At the end we study the distributions of the persistent
regions and the intervals between these. The region size distributions
depend on the initial cluster size distribution. The interval size
distributions are universal, simple exponentials, and their scaling
nicely demonstrates the effect of the presence of the two length
scales, $\Lp$ and $\LD$. 

This paper starts by introducing the model and describing the
quantities of interest in section~\ref{model}. The 
known results for the persistence probabilities 
are presented in the beginning of
section~\ref{sec_Persistence_probabilities}. In the rest of that
section each persistence probability is 
analyzed separately. Section~\ref{persintandregiondists} discusses the
scaling of the 
region and interval size distributions. The dependence of the
persistent quantities on concentration and initial conditions are
studied by simulations in the beginning of
section~\ref{simulations}. The end of 
that section shows the numerical results for the region and interval
size distributions. Section~\ref{conclusions} concludes the paper.

\narrowtext
\begin{figure}
\centering
\includegraphics[width=\linewidth%,draft
]{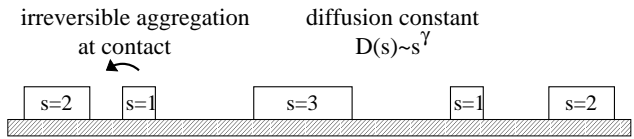}
\caption{Illustration of the one-dimensional DLCA model.  
} \label{DLCAfigeps}
\end{figure}

\section{model and quantities of interest}
\label{model}

The diffusion--limited cluster--cluster aggregation
model is here 
considered on a one-dimensional lattice of $L$ sites with periodic
boundary conditions. Initially the
lattice is filled upto a concentration $\phi$ such that occupied
lattice sites correspond to particles and sites connected via nearest
neighbor occupancy belong to the same cluster. Each cluster performs a
random walk such that the diffusion coefficient of a cluster of size $s$
is given by $D(s) = D_1 s^\gamma$. The positive constant $D_1$ sets
the time scale and it 
is irrelevant regarding the dynamic scaling properties of the system
which are controlled by the diffusion exponent
$\gamma$~\cite{vanDongen:PRL54,Kang:PRA33,Miyazima:PRA36,Hellen:PRE62}.
Here we concentrate on $\gamma < 2$ for which the growth of the
average cluster size is algebraic. For $\gamma >2$ the system will
gel, \ie, there will be an infinite cluster at a finite time. When 
two clusters collide, they aggregate irreversibly
together. If the sizes of the colliding clusters were $s$ and $s'$,
the aggregate is of size $s+s'$ and its diffusion
constant $D(s+s')$. Figure~\ref{DLCAfigeps} visualizes the model.

The one-dimensional DLCA is closely related to other models when the
diffusion coefficient of a cluster is independent of its size, \ie,
$\gamma = 0$. For example, consider the zero temperature
one-dimensional $q$-state Potts model with Glauber dynamics. The
different spin species are separated by domain walls, each of which
performs a simple random walk. By representing each such a walker by a
particle $A$, the two particles colliding react according to  
\begin{equation}
A+A \to 
\begin{cases}
A         & \mbox{with probability} \ \ \ (q-2)/(q-1)  \\
\emptyset & \mbox{with probability} \ \ \ 1/(q-1).
\end{cases} \label{PottsRWeqs}
\end{equation}
In the limit $q \to \infty$ this reduces to the reaction--diffusion
model $A + A \to A$, which can be solved
exactly~\cite{ben-Avraham_in_Privmans_book}. The same is true for
particles with a finite mass $A_i + A_j \to
A_{i+j}$~\cite{Spouge:PRL60}. The only difference between this and 
the DLCA model is the finite extent of clusters in the DLCA. It has no
relevance on the cluster size distribution but the difference may be
essential for persistence. This is to be expected since i) the
$q$-state Potts model can be mapped to the Ising model ($q=2$) with
an initial fraction $1/q$ of up spins~\cite{Sire:PRE52} and ii) the  
persistence exponent of the Ising model is known to depend on
the concentration~\cite{Krapivsky:PRE56,Ben-Naim:JSP93}. Note, that
in the Potts model the usual definition for persistence characterizes
the fraction of spins, which have remained in their original state.

There are many ways to define persistence in cluster--cluster
aggregation. Perhaps the most closest to the definition used in the
context of Ising systems would be the one which relates to the
question of site persistence: what is the probability, that a site has
been either covered or uncovered 
by a cluster for all times $[0,t]$? However, in DLCA there are also
other natural candidates.

\narrowtext
\begin{figure}
\centering
\includegraphics[angle=-90,width=\linewidth%,draft
]{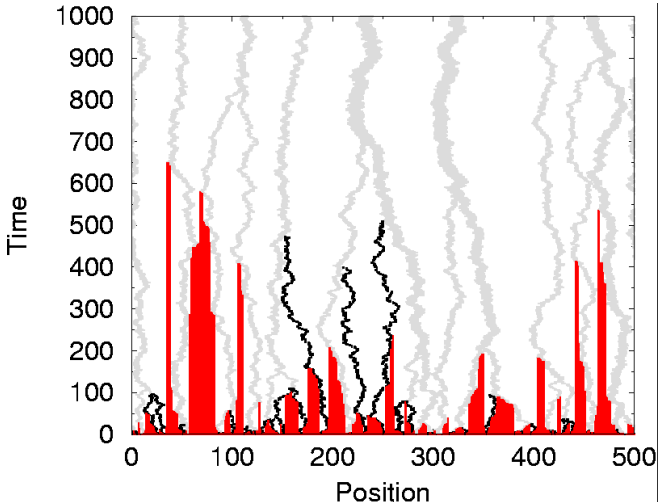}
\caption{A space--time plot of the dynamics in DLCA for $\gamma =
0$. The clusters are denoted by light gray, the persistent clusters by black
and the persistent empty regions by dark grey. 
} \label{spacetime}
\end{figure}

The persistence probabilities studied in this work are the
following:
\begin{itemize}
\item Cluster persistence: 
the probability that a cluster has not aggregated,
$P_C(t) \sim t^{-\theta_C}$. 
\item Filled site persistence: 
the probability that a site originally covered by a cluster has
been covered by it all the time,
$P_F(t) \sim t^{-\theta_F}$. 
\item Empty site persistence: 
the probability that an originally empty site has never been occupied
by a cluster,
$P_E(t) \sim t^{-\theta_E}$.
\end{itemize}
When the persistence probabilities decay algebraically, the
associated decay exponents $\theta_C$, $\theta_F$, and $\theta_E$
become persistence exponents. 
Figure~\ref{spacetime} shows an example 
of the dynamics together with the persistent clusters and the
persistent empty sites for a small system. 

In the following we will concentrate only on the three
definitions given above. However, some other persistences can be
directly obtained from the results given. For example,
the aforementioned site persistence probability is defined as  
the probability that a site has remained in the same state, \ie\
either filled or empty, all the time and is given by $P_S(t) = P_F(t)
+ P_E(t)$. The survival of domains follows from the studies of the
persistent region size distribution (defined below). This is a special
case of the so called adaptive persistence~\cite{Majumdar:PRE57}.

Furthermore one could consider the particle persistence,
which is the probability that a site has remained either empty or
covered by {\it the same\/} particle all the time. Recall that a
cluster of size $s$ consist of $s$ particles.
The probability that a cluster has not
moved before time $t$ decays exponentially making particle persistence
a rather trivial quantity.

The cluster persistence differs drastically from the other persistence
definitions. It is {\it not\/} a quantity defined per
a fixed site on the lattice as the others are but is a property
associated with each cluster. Due to that also its behavior is rather
different than that of the other persistences considered. However, it is
a rather similar, though not exactly the same, quantity as the domain
persistence in the Ising
model~\cite{Krapivsky:PRE56,Ben-Naim:JSP93}. For point-like clusters
it leads to a study of survival of coalescing random walks. A
similar study has very recently been carried out for non-interacting
walks and for the reaction-diffusion system $A+B \to \emptyset$ 
in~\cite{ODonoghue:CM1} and for the annihilation-coalescence
reaction-diffusion system corresponding to the Potts model
[Eq.\eqref{PottsRWeqs}] in~\cite{ODonoghue:CM2}.

In addition to the persistence probabilities we will also study
distributions associated with the persistent sites 
and the distances between them. The terminology and notation
used are as follows. The word region is reserved for a bunch of
consecutive persistent 
sites. The distances between regions, \ie, between two consecutive
persistent sites, is called an interval. The word cluster has the
obvious meaning. The number of clusters of size $s$ per lattice
site at time $t$ is $n_s(t)$ with the normalization $\sum_s s n_s=1$. 
Region size is denoted by $l$ and the number of regions of size $l$
(per site) by 
$p_X(l;t)$. The subscript is like for the persistence probabilities
and it refers to the persistence definition used: $X \in
\{C,F,E\}$. When using the continuum description we use the symbol 
$r$ instead of $l$. The letter $k$ labels the interval sizes and the 
corresponding distribution function is $n_X(k;t)$. The number densities
of persistent sites (the persistence probability) and
nonpersistent ones are denoted by capital 
letters $P_X(t)$ and $N_X(t)$, respectively. 

As an example, consider intervals between persistent empty sites and
their distribution function 
$n_E(k;t)$. The corresponding distribution of persistent regions is
$p_E(l;t)$. The number densities are obtained by summing
\begin{eqnarray}
N_E(t) &=& \sum_{k=1}^\infty \ k \; n_E(k;t) \label{sumNXt} \\
P_E(t) &=& \sum_{l=1}^\infty \ l \; p_E(l;t) \label{sumMXt}. 
\end{eqnarray}
Obviously these two are related by the equation
$P_E(t) = 1 - N_E(t)$. 
Similar formulas apply to other persistence definitions, too, except
in the cluster persistence case 
equation~\eqref{sumMXt} is replaced by
\begin{equation}
P_C(t) = \sum_{l=1}^\infty  p_C(l;t)
\end{equation}
and naturally $P_C(t) \neq 1 - N_C(t)$.

\section{Persistence probabilities} \label{sec_Persistence_probabilities}

In this section we first list the known exact results 
for $\gamma =0$. To discuss the cluster
persistence we use the known results of the dynamic scaling of the
cluster size distribution. We also briefly comment on how the cluster
persistence exponent is related to the polydispersity exponent, which
characterizes the behavior of the scaling function of the cluster
size distribution. Then we apply mean-field type continuum
methods to analyze the empty site persistences for $\gamma
\neq 0$ where no exact results are available. The filled site
persistence is analyzed in a similar mean-field fashion in the limits
$\phi \to 0$ and $\phi \to 1$. This section ends to a discussion on
the concentration dependence of the filled site persistence.

\subsection{Known Exact Results} \label{analytics}

For the cluster persistence the finite extent of the clusters does not
matter and we may replace them by point-like random walkers. 
Then we can utilize the results obtained for the the
reaction-diffusion model $A_i + A_j \to A_{i+j}$.
For spatially
uncorrelated initial mass distribution 
$n_s(0) = \delta_{s,1}$ the cluster size distribution is given by $n_s(t)
= 2se^{-\xi}I_s(\xi)/\xi$, where $\xi = 4D_1t$ and $I_s(\xi)$ is the
modified Bessel function~\cite{Spouge:PRL60}. As $n_1(t) \sim
t^{-3/2}$ the cluster 
persistence exponent $\theta_C(\gamma=0) = 3/2$ in this case. 
This result is also valid independent of the initial condition as long
as the cluster size distribution decays rapidly for large sizes. 

The result $\theta_C(0) = 3/2$ can also be obtained without solving
the cluster size 
distribution for general initial size distribution of
clusters. Consider a cluster initially at the origin. We want to 
find the probability that it has not collided with neither of
its neighboring clusters before time $t$. As the diffusion
coefficient of a cluster does not depend on its size, we do not have to
care if the neighboring clusters have collided with other clusters
or not. Therefore the problem reduces to the survival probability of
three annihilating random walkers. This can 
solved in several
ways~\cite{Fisher:JSP34,Derrida:PRE54} with the result
$\theta_C(\gamma=0) = 3/2$. Unfortunately, these methods rely on  
the indistinguishability of clusters and can not simply be generalized
to the general case of size dependent diffusion coefficients. In fact,
even the survival of three annihilating random walkers with constant but
non-equal diffusion coefficients is highly nontrivial~\cite{Fisher:JSP53}.
In DLCA it is the specific interaction
between clusters, the aggregation at contact, that makes the problem
simple. The case 
where a single cluster is hunted by independent clusters is in fact
more complicated than our situation with aggregating clusters (see
e.g.~\cite{Krapivsky:JPA29}).

Let us now turn to site persistence when $\gamma = 0$. For the empty
site persistence 
the finite extent of the clusters does not matter since a persistent
interval is deleted by the edges of neighboring clusters. For finite
size clusters the empty intervals are shorter but the length
distribution remains the same. The movement
of the edges is uncorrelated and unaffected by the collisions
with other clusters. These facts make the problem exactly solvable
with the result $\theta_E(\gamma=0) = 1$~\cite{Bray:PRE62}. Note, that
although the movement of 
edges is uncorrelated for $\gamma \ne 0$, too, one can no more ignore
the collisions of the neighboring clusters with other clusters. This
makes the exact calculation complicated if not intractable. The same
applies to the filled site persistence even for $\gamma = 0$.

\subsection{Cluster persistence} \label{Clusterpersistencesec}

Before considering cluster persistence we briefly recapitulate the
main results concerning the dynamic scaling in the DLCA. Let the
number of clusters of size $s$ per lattice site at time $t$ be
$n_s(t)$.  The cluster size distribution scales as 
\begin{equation}
n_s(t) = S(t)^{-2} f(s/S(t)),
\end{equation}
where $S(t)$ is the average cluster size and $f(x)$ the scaling
function~\cite{Meakin:PhysicaScripta46}. The diffusive length scale,
$\LD$, is proportional to the average cluster size 
which grows as $S(t) \sim t^z$, where $z$ is the dynamic exponent. For
a given $\gamma$ both $z$ and the scaling function are universal. In
one dimension the dynamic exponent is given by $z = 1/(2-\gamma)$ 
for $\gamma<2$ and the system gels in a finite time for
$\gamma>2$~\cite{Kang:PRA33,Miyazima:PRA36}. For 
small argument values ($x \to 0^+$) the
scaling function $f(x)$ decays as $x^{-\tau}$ for $0 \le \gamma <2$
and as $\exp(-x^{-|\mu|})$ for
$\gamma<0$~\cite{vanDongen:PRL54,Hellen:PRE62}. Though universal, both
the exponents $\tau$ and $\mu$ depend nontrivially on $\gamma$. From the
small $x$ behavior of the scaling function it follows that $n_s(t) \sim
s^{-\tau}t^{-w}$ for $1 \ll s \ll S(t)$, where $w=(2-\tau)z$.

The scaling theory further states that for the DLCA all
the cluster densities decay in a similar manner at large times:
$n_s(t)/n_1(t) \to b_s$ as $t \to \infty$~\cite{vanDongen:JPA18}. Here
$b_s$ is a time independent constant. 
This result together with $n_s(t) \sim t^{-w}$ can be directly applied
to cluster persistence with a monodisperse initial condition $n_s(0) =
\delta_{s,s_{_0}}$. For
simplicity, let all the clusters be monomers, \ie, of unit size, at
time $t=0$. Then the cluster persistence probability is equal to
$n_1(t)\sim t^{-w}$ giving 
\begin{equation}
\theta_C = (2-\tau)z. \label{thetaCscalrel}
\end{equation}
The exact exponents $\tau=-1$, $z=1/2$~\cite{Spouge:PRL60}, and
$w=3/2$~\cite{Fisher:JSP34,Derrida:PRE54} for $\gamma=0$ satisfy this
relation. Simulations show that similarly than $z$ and $\tau$ also the
persistence exponent is independent of initial conditions and
concentration (see Sec.~\ref{subsecDep_in_cond}). 

The result $\theta_C = (2-\tau)z$ deserves a few comments. First,
since the cluster persistence is a cluster property, it is
natural that the persistence exponent is related to the cluster size
distribution. 
Since a persistent cluster has not collided with others, it will
definitely be smaller than the average cluster size at late
times. Thus the persistent clusters are the ones represented by the small
$s$-tail of the cluster size distribution,
which is described by the polydispersity exponent
$\tau$. On the other hand, if there is any scaling relation between the
cluster persistence exponent and the traditional exponents, this
should include the dynamic exponent. On the basis of this
reasoning, the above result is reasonable. However, most of
the scaling results quoted above are mean-field of nature and
therefore Eq.~\eqref{thetaCscalrel} has to be validated by
simulations. The same scaling relation has been found
also in the diffusion-annihilation model discussed in
the Introduction~\cite{Manoj:JPA33}. 

It follows from the scaling relation that the cluster
persistence exponent is as universal as the exponents $z$, $w$,
and $\tau$. However, it does not result in a value for
the persistence exponent since $\tau$ is a priori
unknown. Calculating it for DLCA and 
other similar processes has turned out to be a
challenging problem since it is given in terms of integrals over the
scaling function $f(x)$ itself~\cite{vanDongen:PRL54,Cueille:PRE55}. 
Also determining the value of $\tau$ reliably in simulations is hard.
On the other hand, if one could calculate the persistence
exponent and validate the scaling relation one would also obtain $\tau$. 

In fact, the persistence exponent can be calculated in a mean-field
fashion by replacing the clusters surrounding a persistent cluster by
random walkers which diffuse as an average cluster would
diffuse.  
A similar approach works extremely well also for the empty site
persistence (see Sec.~\ref{emptysiteperssec}). Here it
results in studying the survival of three random walkers, whose
diffusion coefficients are time-dependent. The 
meaning of a time-dependent diffusion coefficient in this context is
that a random walker will diffuse with a constant diffusion
coefficient in another time scale. The detailed analysis and the
numerical verification of Eq.~\eqref{thetaCscalrel}
will be presented elsewhere~\cite{omapitka} and we just give the
result here. 

It turns out that the cluster persistence probability 
\begin{equation}
P_C(t)  \sim
\begin{cases}
\exp(- C t^{\beta_S})  &, \gamma < 0 \\
t^{-3/2}  &, \gamma = 0 \\
t^{-2/(2-\gamma)}  &, 0 < \gamma < 2 \label{cpreseq}
\end{cases}
\end{equation}
with $C>0$ a constant. $\beta_S$ fits well to the expression
$\beta_S = 2/3(1-2z)$.  
Here the completely different behavior of the cluster
persistence for $\gamma<0$ and $\gamma \ge 0$ is similar to that of 
the cluster size distribution which changes character at $\gamma=0$
as noted in the beginning of this Section. The persistence exponent is
discontinuous as $\gamma \to 0^+$. 
Quite unexpectedly, since
$\theta_C(0<\gamma<\frac{2}{3}) < \theta_C(0)$,
it is easier for a cluster to remain persistent if the other clusters
will diffuse faster! This somewhat paradoxical result
can be understood on the basis that for a random
walker, which is hunted by other random walkers, the optimal strategy
to survive is to remain stationary (see~\cite{Redner:AJP67} for a similar
situation in the case of noninteracting walkers hunting a single
random walker). In the limit $\gamma \rightarrow \infty$ the 
stretching exponent $\beta_S \rightarrow 2/3$ is also ``discontinuous'', since
the persistence for a particle bounded by two immobile neighbors
decays simply exponentially.

The result~\eqref{cpreseq} together with Eq.~\eqref{thetaCscalrel}
implies that also the polydispersity exponent is discontinuous at
$\gamma = 0$ and, more interestingly, that $\tau(\gamma) = 0$ for
$0<\gamma<2$. Although this kind of a ``superuniversality'' is
remarkable in its own right, it is more important, that by studying
the cluster persistence one learns something about the polydispersity
exponent, \ie, the scaling function.
This connection may exist in other aggregation models 
offering a new way to approach the problem of computing the
small size tail of the scaling function through a perhaps simpler
quantity, the cluster persistence.

\subsection{Empty site persistence} \label{emptysiteperssec}

Next we give a heuristic argument for obtaining the empty site
persistence exponent for arbitrary $\gamma$. Since the 
clusters on both sides of a persistent empty region are
independent, we are led to consider the maximum excursion of a single random
walker. The only complication comes from the fact that for
$\gamma \neq 0$ the collisions of a cluster with other clusters will
change its diffusivity. 

Instead of the real process we consider 
each cluster to have a {\it time-dependent} diffusion coefficient $D(t) =
D_0 t^{\gamma z} \sim S(t)^\gamma$ with $z=1/(2-\gamma)$. Then the
probability, $P(x;t)$, to find a cluster at position $x$ at time $t$ obeys a
diffusion equation 
\begin{equation}
\partial_tP(x;t) = D_0 t^{\gamma z} \label{seceqdif}
\partial_x^2P(x;t).
\end{equation}
A time transformation
\begin{equation}
T(t) = \frac{D_0}{\gamma z+1} t^{\gamma z+1} 
\label{Ttequation} 
\end{equation}
reduces equation~\eqref{seceqdif} to an ordinary diffusion equation
with the diffusion constant $D=1$. The persistence of the empty
space between particles diffusing 
with a constant $D$ has recently been considered in~\cite{Bray:PRE62} and we
just quote the main results here. 

In the long time limit the probability density, $p_E(r;t|s)$, that a
persistent empty region (originally of size $s$) has size $r$ at
time $T$ is given by
\begin{equation}
p_E(r;T|s) = \frac{s-r}{\pi T} \label{PlTl0eq}
\end{equation}
and the probability that a cluster survives upto time $t$ is
\begin{equation}
P_E(t|s) = \frac{s^2}{2\pi T} = \frac{z}{\pi D_0 }
\frac{s^2}{t^{2z}} \sim t^{-2z}.
\end{equation}
For a general initial length distribution of clusters $n_s(0)$ the
result will remain the same except $s^2$ will be replaced by the
average over the initial
length distribution $\langle s^2 \rangle$. 

These considerations show that the persistence exponent 
$\theta_E(\gamma) = 2z = 2/(2-\gamma)$. This agrees with the exact
result $\theta_E(0) =  1$. Furthermore the result is independent of
the initial spatial distribution or concentration. All these
implications based on the simplified process are confirmed by
simulations (see Secs.~\ref{sim_prob}-\ref{depongammasec}).

\vspace{5cm}

\narrowtext
\begin{figure}
\centering
\includegraphics[width=\linewidth%,draft
]{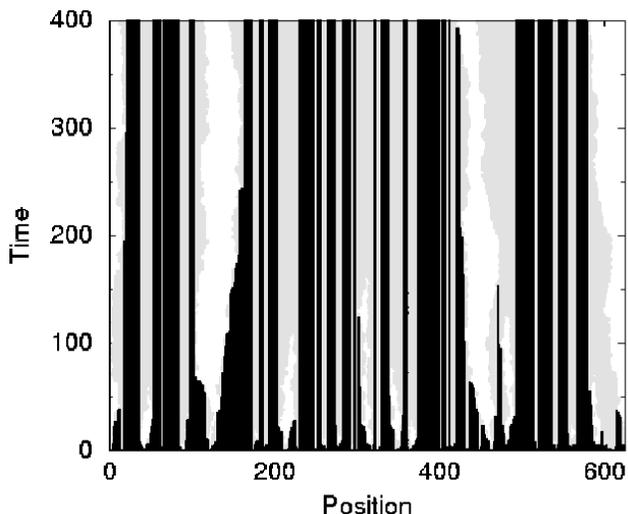}
\caption{A space--time plot for concentration $\phi=0.8$ and $\gamma
=0$. Cluster are denoted by grey and the persistent empty sites by black.
} \label{largecons_pers_filled} 
\end{figure}

\subsection{Filled site persistence} \label{subsecFilledsp}

The behavior of the filled site persistence is 
different at low and high concentrations. At low concentration a
cluster contains typically only one persistent region which
is usually destroyed before the cluster aggregates with a
neighbor. At high concentrations
a large cluster has many persistent regions of various sizes inside
it due to the aggregation of clusters (Fig.~\ref{largecons_pers_filled}). 
We now analyze these two limits separately.

In the low concentration limit we
consider the persistence problem in continuum, for simplicity. This
should be valid for clusters of initial length $l_0 \gg 1$. 
For $\phi \to 0$ the time required for a cluster to move its
own length is much smaller than the time required to overcome the
distance between clusters. Therefore one could assume that at low
concentrations collisions between clusters do not matter.
We will see in section~\ref{sim_prob} that this
is true only upto some crossover-time.
This time obviously diverges in the limit 
$\phi \to 0$. However, to obtain more insight to the empty site
persistence we will first ignore the collisions in the following analysis.

\narrowtext
\begin{figure}
\centering
\includegraphics[width=\linewidth%,draft
]{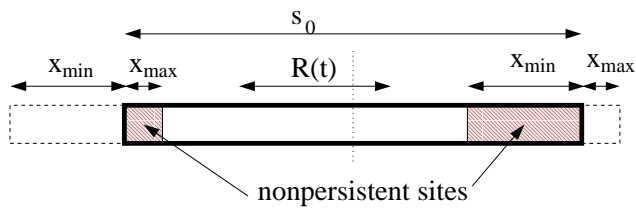}
\caption{The length of the nonpersistent filled part is equal to the
span $R(t) = x_{\rm min}(t) + x_{\rm max}(t)$ of the middle point of a
cluster. The length of the persistent region is $s_0 - R(t)$. The
cluster position at $t=0$ is denoted by a thick 
rectangle, whose middle position is marked by a dotted line. The
dashed lines show the maximum excursions of the cluster. 
} \label{spanfig} 
\end{figure}

When clusters do not collide, the persistent sites under 
different clusters are destroyed by independent processes. 
A single diffusing cluster destroys persistent area at both of its 
edges. As figure~\ref{spanfig} shows, the length of the nonpersistent filled
part inside an unaggregated cluster is equal to the span of the middle
point. The span, $R(t)$, of a random walk is
defined as $R(t) = x_{\rm min}(t) + 
x_{\rm max}(t)$, where $x_{\rm min}(t)$ and $x_{\rm max}(t)$ are the
maximum displacements of the walk in the negative and
positive directions at time $t$, respectively.

The probability distribution for the span of an unlimited random walk
is given by~\cite{Weiss:book} 
\begin{equation}
w(R;t) = \frac{8}{\sqrt{2\pi D t}}\sum_{j=1}^\infty (-1)^{j+1} j^2
\exp \left(-\frac{j^2R^2}{2Dt} \right). \label{spaneq}
\end{equation}
In our case the maximum span is limited by the size of the cluster, $s$,
and the probability distribution $p_F(r;t|s)$ that an interval of length
$r$ of a cluster initially of size $s$ has survived upto time $t$ is
\begin{equation}
p_F(r;t|s) = \frac{8}{\sqrt{2\pi D t}}\sum_{j=1}^\infty (-1)^{j+1} j^2
\exp \left(-\frac{j^2(s-r)^2}{2Dt} \right) \label{u(r;t|s)}
\end{equation}
for $r\le s$ and zero otherwise. 

We are interested in the asymptotic long time behavior and 
rather than working directly with the formula in Eq.~\eqref{u(r;t|s)}
it is useful to transform it to a more tractable form as
follows. Writing the sum in Eq.~\eqref{u(r;t|s)} in the form
$\sum_{j=-\infty}^{\infty}G_j$ and applying the Poisson sum
formula~\cite{Kadanoff:book}
\begin{equation}
\sum_{j=-\infty}^{\infty}G_j = \sum_{m=-\infty}^{\infty}
\int_{-\infty}^\infty {\rm d}x\ G(x) e^{2\pi i m x}, \label{PoissonSum}
\end{equation}
leads to
\begin{eqnarray}
p_F(r;t|s) 
&=& \frac{8 D t}{(s-r)^3} \sum_{m=0}^{\infty} \left[
\frac{\pi^2Dt(2m+1)^2}{(s-r)^2} -1\right]  \nonumber \\ 
&&\times \exp
\left(-\frac{\pi^2Dt(2m+1)^2}{2(s-r)^2} \right). \label{handyform}
\end{eqnarray}
In Eq.~\eqref{PoissonSum} $G(x)$ is an analytic function having the
same values at all integers as $G_j$. 

From Eq.~\eqref{handyform} it is easy to show that the probability of
finding persistent sites inside the cluster decays exponentially at
large times:
\begin{eqnarray}
p_F(t|s) &=& \int_0^s {\rm d}r\ p_F(r;t|s) \nonumber \\
&=& \sum_{m=0}^{\infty} 
\frac{8[s^2+\pi^2Dt(2m+1)]}{s^2\pi^2(2m+1)^2}  \nonumber \\
&& \times \exp \left(-\frac{\pi^2Dt(2m+1)^2}{2s^2} \right)  \\
&\approx& 
\frac{8Dt}{s^2} \exp \left(-\frac{\pi^2Dt}{2s^2} \right). \label{apptildeutl}
\end{eqnarray}
The persistence probability is obtained by integration 
\begin{equation}
p_F(t) = \int_0^\infty {\rm d}s\ p_F(t|s) n_s(0),
\end{equation}
so that, in general, the decay will depend on $n_s(0)$. 

For $n_s(0) = \delta(s-s_0)$ it is given by an exponential expression
[Eq.~\eqref{apptildeutl}] but 
the change in the decay can be shown using, for example, the initial
distribution 
\begin{equation}
n_s(0) = \frac{2}{\pi s_0} \exp\left(
-\frac{s^2}{\pi s_0^2} \right),
\end{equation}
whose mean and standard deviation are 
$\langle s \rangle = s_0$ and $\sigma_s = \sqrt{\langle s^2
\rangle -\langle s \rangle^2} = s_0\sqrt{(\pi-2)/2}$, respectively. 
Using the
approximation~\eqref{apptildeutl} we get a stretched exponential
\begin{equation}
p_F(t) \approx 
\sqrt{\frac{128Dt}{\pi^3s_0^2}} 
\ e^{-\sqrt{2\pi D t/s_0^2}}.
\end{equation}
For the Poisson initial distribution, $n_s(0) =
s_0^{-1}e^{-s/s_0}$, we can apply the saddle point method to give
\begin{equation}
p_F(t) \approx \sqrt{\frac{128Dt}{3\pi s_0^2}}
\exp\left( -\frac{3}{2} \left(\frac{\pi^2 Dt}{s_0^2}\right)^{1/3}
\right). \label{uFtsaddlepoint}
\end{equation}

Both of these examples show a stretched exponential decay for the
persistence probability, which we believe to be true quite
generally. The stretching exponent depends on the initial
condition and is therefore nonuniversal. 
The long time behavior, in the approximation ignoring collisions, is
governed by the ratio $Dt/s_0^2$ and we also
immediately see that the decay
depends on concentration due to the factor $s_0$ in the exponential. 

The approximative span distribution can be also used to calculate 
the mean size of persistent regions.
First write
\begin{equation}
\langle r(t|s) \rangle = \langle s-R(t|s) \rangle = s - \langle R(t|s)
\rangle 
\end{equation}
and use the normalized approximate form for the span distribution 
\begin{equation}
\tilde{w}_N(R;t|s) = 
\begin{cases}
 \delta(R-s) 
 \left[ 1-\left(1+\xi s^{-2}\right) e^{-\xi/s^2} \right] & \\
\ + \ 2\xi^2R^{-5} e^{-\xi/R^2}  &; R \le s\\
 0 &; R>s,
\end{cases}
\end{equation}
where $\xi = \pi^2Dt/2$,
to calculate $\langle R(t|s) \rangle$ from which
\begin{equation}
\langle r(t|s) \rangle 
\approx s \exp\left\{ -\frac{\pi^2Dt}{2s^2} \right\}.
\end{equation}
For the simplest case of a fixed initial size $s_0$  [$n_s(0) =
\delta(s-s_0)$] the mean length of surviving regions is 
\begin{eqnarray}
\langle s_{\rm surv} \rangle = \frac{1}{p_F(t|s_0)}
\int_0^{s_0} {\rm d}r\ rp_F(r;t|s_0) \approx \frac{2s_0^3}{\pi^2 Dt} \sim
t^{-1}. 
\end{eqnarray}
This is just an example and naturally the decay depends on the
initial distribution $n_s(0)$.

In the high concentration limit, $\phi \to 1$, we 
adopt another mean-field type approach: we consider a deterministic model
combined with scaling arguments. Let initially 
the average cluster and empty interval size be $s_0$ and  size $d_0$,
respectively. The concentration $\phi = s_0/(s_0+d_0) \approx 1$ for $s_0
\gg d_0$.  

Now consider the doubling times $t_1, t_2, \ldots, t_i, \ldots$ at which
the average cluster and interval sizes are 
$s_i = 2^is_0$ and $d_i = 2^id_0$,
respectively. At each step $n$ the doubled cluster is
constructed as follows. First, $d_{n-1}/2$ sites (these does
not have to persistent but they may be) from both ends of the cluster
are made nonpersistent. Second, the resulting cluster is
duplicated. This procedure is repeated. It is quite easy to convince
oneself that the probability of finding persistent sites, $p(n)$, at
step $n$ decays 
like $p(n) \sim e^{-\alpha(\phi)n}$ for high enough values of
$\phi$.
Since $s_n = 2^ns_0 \sim t^z$,  
it follows that $p(t) \sim t^{-\alpha(\phi)z/\ln 2}$.

According to this simple and crude argument the filled site
persistence probability decays algebraically for  
large enough concentrations. 
The persistent sites are sweeped by the domain walls (clusters'
edges), which get destroyed at aggregation. 
Since the probability to be touched by a domain wall depends
naturally on the density of the walls in the system,
the persistence exponent depends on
concentration implying nonuniversality. Simulations
qualitatively agree with this behavior. We have not made any
quantitative comparison since in this paper we concentrate more on the
universal properties of the cluster--cluster aggregation and since
being non-universal the decay of site persistence is harder to
compute analytically. The
simulations furthermore show that the persistence probability decays
algebraically for low concentrations, too. The reason for the 
deviation from the span argument lies in
the approximation, which neglects the collisions between clusters. We
return to this issue in a more detail in the section~\ref{sim_prob}.

\section{Persistent interval and region distributions} \label{persintandregiondists}

Having analyzed the various persistence quantities we now turn to the
distributions of persistent regions and the intervals between them.
First we consider the interval size distributions between consecutive
persistent sites and perform a scaling analysis.  
A natural scaling assumption for these distributions is 
\begin{equation}
n_X(k;t) = K_X(t)^{-\alpha} f_X(k/K_X(t)), \label{nXktscalform}
\end{equation}
where $K_X(t) \sim t^{\theta_X}$ is the average interval size and $X$
denotes any of the choices $C,E$, or $F$. 

Inserting the scaling form~\eqref{nXktscalform} to the
equation~\eqref{sumNXt} and replacing the sum by 
an integral (valid in the long time limit) results in 
\begin{equation}
N_X(t) = K_X(t)^{2-\alpha} \int_0^\infty {\rm d}y \;
yf_X(y). \label{NXinteq} 
\end{equation}
As $P_X(t) \to 0$ for $t \to 
\infty$ it follows from the relation $P_X(t) = 1 - N_X(t)$ that
$N_X(t) \to 1$ as $t \to \infty$. 
The only way to keep the right hand side of Eq.~\eqref{NXinteq}
constant is to require $\alpha = 2$, as a direct consequence of mass
conservation.
Note, that the above argument does not require the persistence
probability $P_X(t)$ to decay algebraically. 

Since the persistence exponents are larger than the dynamic exponent,
$\theta_X>z$, the persistence length scale will be much larger than
the diffusive one at large times. This means that  
the persistent regions are well separated and that they are destroyed
by uncorrelated processes since the correlations grow only as $t^z$. 
Each persistent region survives independently of the others and in
the scaling limit the scaling functions are therefore simple
exponentials 
\begin{equation}
n_X(k;t) = K_X(t)^{-2} 
e^{-k/K_X(t)}. \label{xXktsimpleexpscal}
\end{equation}

Next let us turn to the size distribution of persistent regions.
As the filled site persistence shows
nonuniversal behavior, we will here concentrate only on the
distributions concerning the empty site and cluster persistence.
The former has already been analyzed in
section~\ref{sec_Persistence_probabilities}.
For example,
equation~\eqref{PlTl0eq} gives the persistent empty region
distribution at large 
times for a monodisperse initial distribution and for other initial
distributions $n_s(0)$ it is obtained as  
\begin{equation}
p_E(r;T) = \int_0^\infty {\rm d}s\ p_E(r;T|s) n_s(0) = \frac{s_0}{\pi T}
e^{-r/s_0}, \label{Eregiondelta}
\end{equation}
where $T$ is given by equation~\eqref{Ttequation} and the last form
corresponds to the initial distribution
$n_s(0)=s_0^{-1}e^{-s/s_0}$. The dependence on the diffusion exponent
enters only through the time scale $T \sim t^{2/(2-\gamma)}$. 

More important is that the spatial and time dependence in $p_E(r;t)$
are decoupled. From this it follows that the average size of the
persistent empty regions, $L_E(t)$, is a constant at large
times. For the monodisperse and
Poisson initial conditions we get $L_E^{\rm mono}(t) = 
P_E(T|s_0)^{-1} \int_0^{s_0} {\rm d}r\ \! r p_E(r;T|s_0) =
s_0/3$ and $L_E^{\rm rand}(t) = 
\int_0^{\infty} {\rm d}r\ \! r p_E(r;T)/\int_0^{\infty} {\rm d}r\ \!
p_E(r;T) = s_0$, 
respectively. These both are independent of time.
The simulations (see Sec.~\ref{sim_regions}) confirm this.

Unfortunately we have not been able to similarly calculate the size
distribution of persistent clusters, $p_C(l;t)$. However, 
for initial cluster size distribution $n_s(0) = \delta_{s_{_0},s}$ this
distribution remains unaltered. The same is true also 
for size independent diffusion coefficients
($\gamma = 0$) no matter what the initial distribution is. This does
not have to be true when $\gamma \neq 0$. For instance, there may be
nontrivial correlations between a persistent cluster and the sizes of
clusters surrounding it. 
The scaling of $p_C(l;t)$ is briefly discussed in
section~\ref{sim_regions}.

\section{Simulations} \label{simulations}

The simulations are done on a one-dimensional lattice with periodic
boundary conditions. In all the simulations $D_1=1$ fixing the time scale.
The system sizes
range from $5 \cdot 10^4$ to $1.5 \cdot 10^6$, the data are averaged over 
$1000-50000$ realizations. 
The algorithm used is as follows. A cluster is selected randomly and the
time is incremented 
by $N(t)^{-1}D_{\rm max}^{-1}$, where $N(t)$  
is the number of clusters at 
time~$t$ and $D_{\rm max}$ is the maximum diffusivity of any of the
clusters in the system at that time. 
If $x<D(s)/D_{\rm max}$, where $x$ is an
uniformly distributed random number in the interval~$[0,1]$, the
cluster is moved with equal probability either to the left or to 
the right. Otherwise a new cluster is selected and the
process is repeated. If a moved cluster collides with another one, the
two clusters are aggregated together.

We test the sensitivity of the 
persistence probabilities against concentration changes
(Sec.~\ref{sim_prob}) and two 
different initial conditions (Sec.~\ref{subsecDep_in_cond}).  
The first initial condition used is random, which means that   
each lattice site is filled with probability $p$. The other one is
deterministic and monodisperse such that clusters of a given size $s_0$
are separated form neighboring clusters by a distance $l_0$ resulting in
the concentration $\phi = s_0/(s_0+l_0)$.

After discussing the universality or nonuniversality of the
persistences we concentrate on the dependence of the persistence
exponents on the diffusion exponent $\gamma$ in
section~\ref{depongammasec}. Thereafter we consider 
the size interval distributions (Sec.~\ref{sim_intervals}). We
demonstrate how the poor dynamic scaling of the interval distributions  
is a consequence of the two competing length scales $\LD$ and
$\Lp$. The scaling function of the region size distributions is shown
to be of the simple exponential form in section~\ref{sim_regions}.

\narrowtext
\begin{figure}
\centering
\includegraphics[width=\linewidth%,draft
]{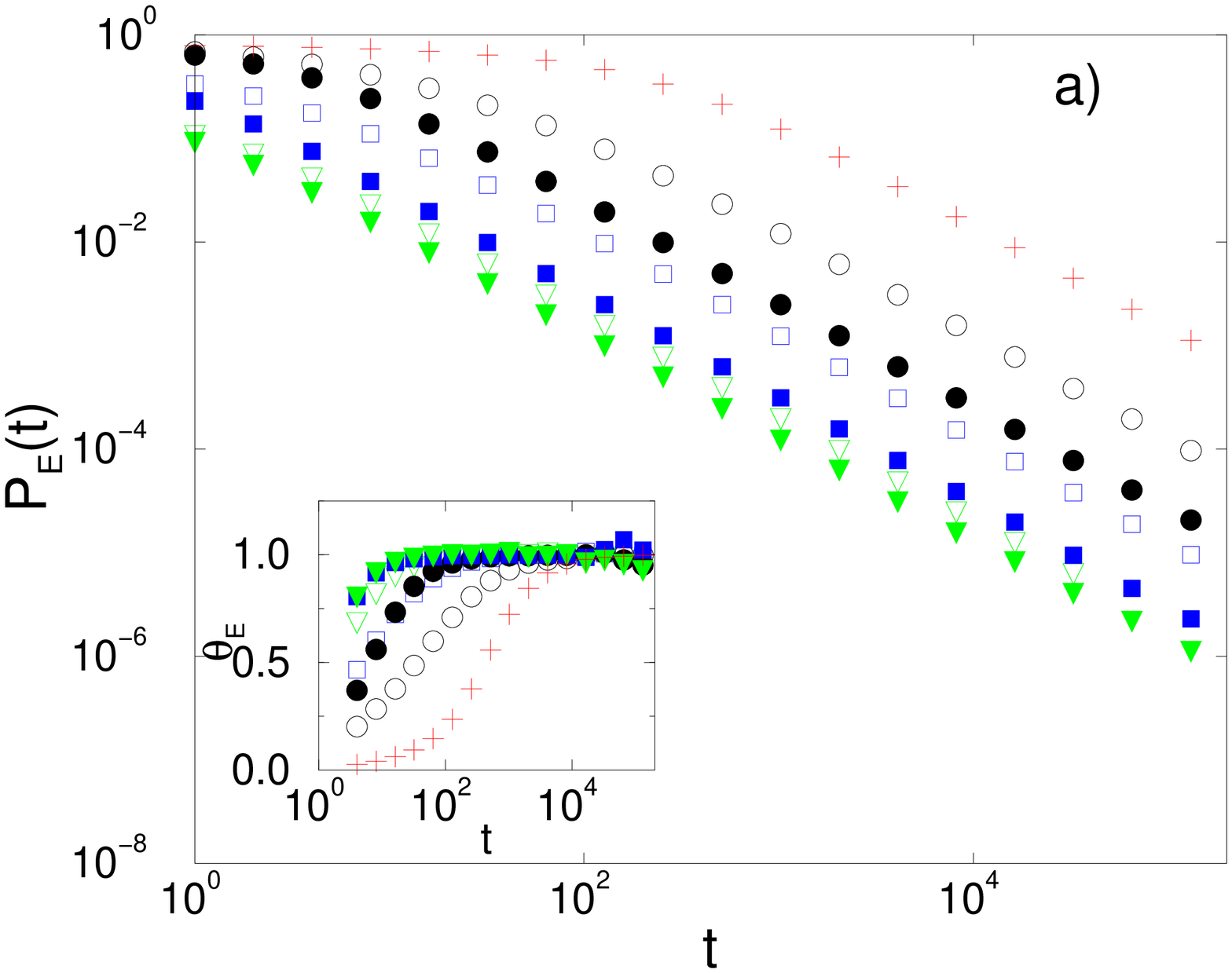}
\includegraphics[width=\linewidth%,draft
]{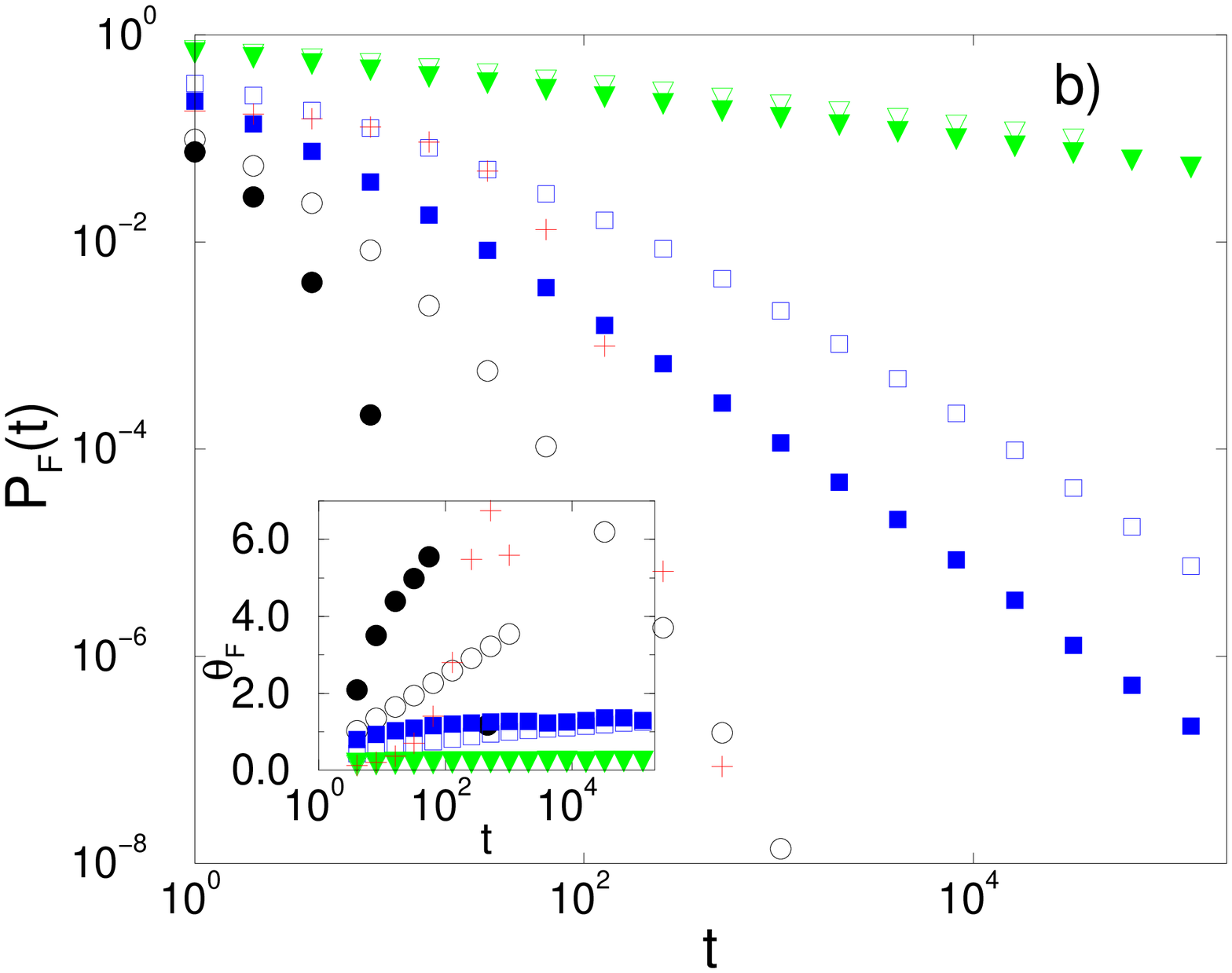}
\includegraphics[width=\linewidth%,draft
]{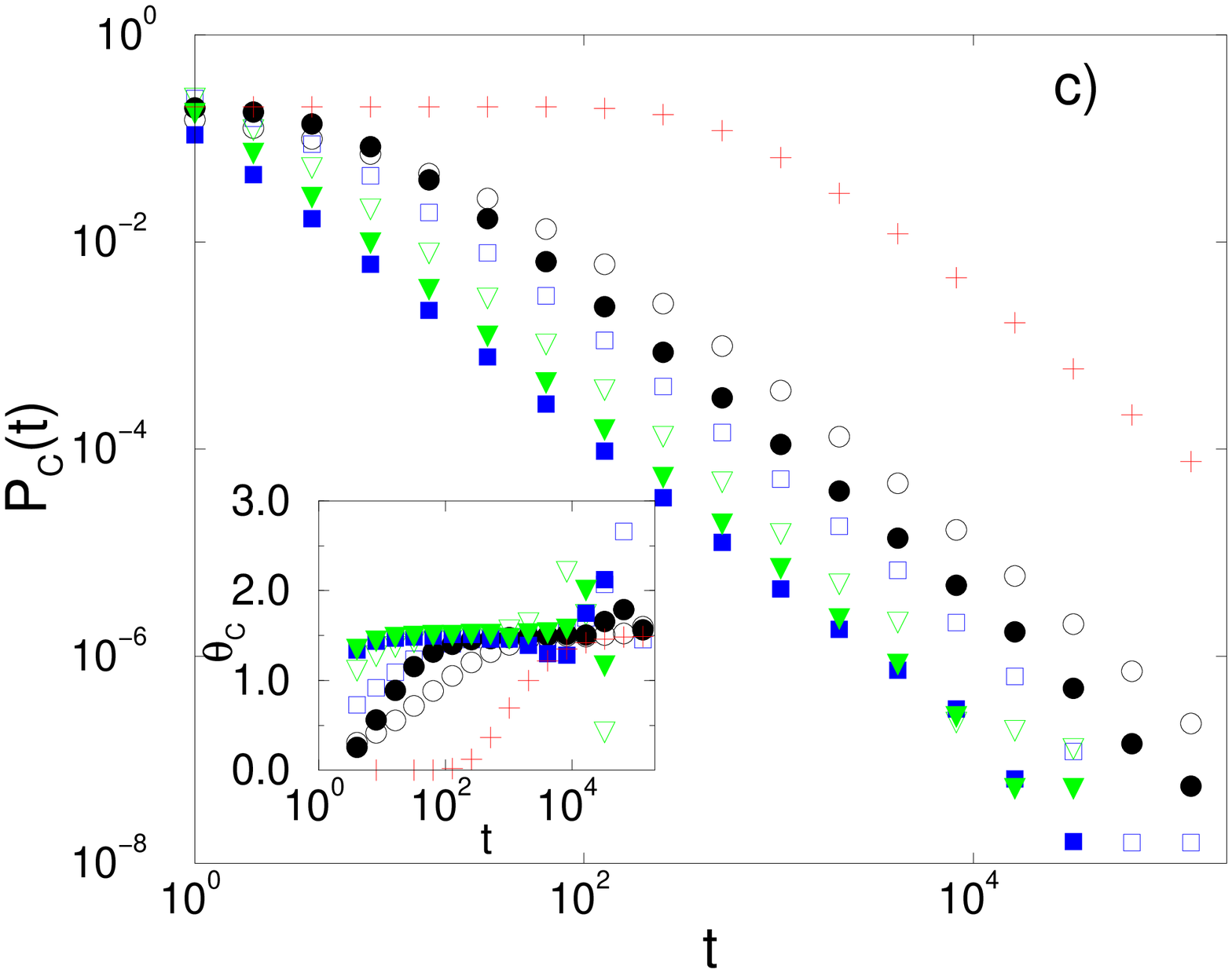}
\caption{The persistence probabilities a) $P_E(t)$, b) $P_F(t)$, and
c) $P_C(t)$ for 
size independent diffusion coefficients ($\gamma = 0$) and for
concentrations $\phi = 0.20$~($\bigcirc$), $0.50$~($\square$), and
$0.80$~($\nabla$). Results from simulations using random and
monodisperse ($s_0=1$) initial conditions are denoted by open and filled
symbols, respectively. Pluses~(+) are obtained with monodisperse
initial conditions for $\phi = 0.20$ and $s_0 = 10$. The insets show
the running exponents. 
} \label{consdepfig} 
\end{figure}

\subsection{Concentration dependence} \label{sim_prob}

We first analyze which persistence probabilities depend on
concentration. We will present data only for size independent
diffusion coefficients but we have checked that for
$\gamma \neq 0$ the persistence probabilities behave in a similar
way. Figure~\ref{consdepfig} shows the cluster, empty  
site, and filled site persistence probabilities for three different
concentrations. A change in the concentration affects only the
amplitudes of the cluster and empty site persistence distributions.
The numerical estimates for the concentration independent persistence
exponents obtained from the saturated part of the running exponents are
$\theta_C(0) = 1.48 \pm 0.03$ and $\theta_E(0) = 1.00 \pm 0.02$. These
both are in an excellent agreement with the exact results $\theta_C(0)
= \frac{3}{2}$ and $\theta_E(0) = 1$.

\narrowtext
\begin{figure}
\centering
\includegraphics[width=\linewidth%,draft
]{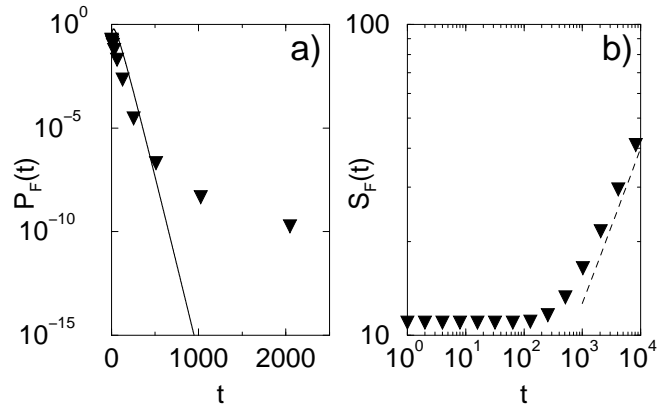}
\caption{a) Filled site persistence probability and b) average cluster
size of those clusters which contain persistent
sites for deterministic initial conditions with $\gamma = 0$, $s_0=11$,
and $\phi = 
0.2$. The solid line is given by equation~\eqref{apptildeutl} and the dashed
line shows $t^{1/2}$ behavior. 
} \label{filledPtheory} 
\end{figure}

For the filled site persistence the effect of a concentration change
is much more pronounced. The distribution shows a transition 
from an algebraic decay to a relatively faster one when
decreasing the concentration. This is in a qualitative agreement with
the analysis of section~\ref{subsecFilledsp} but the agreement
is not quantitative. In fact, the non-algebraic decay seen in
simulations is only a cross-over behavior and eventually the
persistence probability will follow a power-law. 

The discrepancy between the analysis of
Sec.~\ref{subsecFilledsp} and simulations can be traced to the
assumption, that the clusters containing persistent
filled sites have not aggregated. This is true only
for times smaller than the average collision time between clusters,
$t_{\rm coll} \sim 1/(D\phi^2)$, which indeed diverges for $\phi \to
0$. However, for any nonzero concentration there will be a time after
which the collisions become important. In fact, the large time
persistence probability of filled sites is dominated by the
clusters, which have collided with other ones. This is illustrated
in Figure~\ref{filledPtheory}, which shows both the
persistence probability and 
the average size of those clusters which contain persistent
sites. There is a clear crossover from the behavior given by the
analysis of Sec.~\ref{subsecFilledsp} at $t \approx 300$ to the one
for which the collisions are significant. After this crossover time,
the clusters including persistent empty sites grow similarly as the
other ones. This is illustrated in figure~\ref{filledPtheory}~b),
which shows that the average size of these clusters grows like
$t^{1/2}$ at late times.

\subsection{Dependence on initial conditions} \label{subsecDep_in_cond}

Figure~\ref{consdepfig} also compares the persistence probabilities for the 
two initial conditions: random and monodisperse deterministic. A
change in the initial condition does not 
have a significant effect on any of the persistence probabilities. It
only affects the overall amplitude but leaves the qualitative behavior
unaltered.  

Simulations can of course not prove the universality of empty or
cluster persistence but together with the arguments given in
section~\ref{sec_Persistence_probabilities} they strongly support this.
However, they indisputably show that the filled site
persistence is nonuniversal. Although non-universal, the empty
site persistence is closely related to the persistence of bubbles in
soap froths, which show similar concentration dependent
behavior~\cite{Tam:EPL51}. It would be interesting to study the
similarities between these two persistences, but we will omit it
here.

\narrowtext
\begin{figure}
\centering
\includegraphics[width=\linewidth%,draft
]{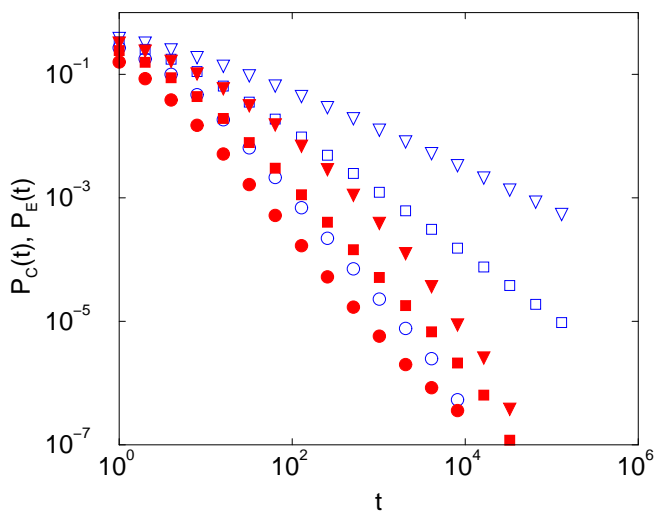}
\caption{The cluster and empty site
persistence probabilities $P_C(t)$ (filled symbols) and  
$P_E(t)$ (empty symbols)
for $\gamma = -1.0~(\nabla), 0.0~(\square)$, and $0.75~(\bigcirc)$.
} \label{pers_prob_gamma_fig} 
\end{figure}

\narrowtext
\begin{figure}
\centering
\includegraphics[width=\linewidth%,draft
]{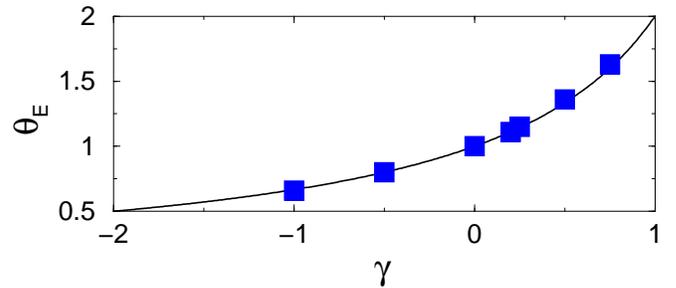}
\caption{Comparison of the numerically obtained empty site persistence
exponent~$\theta_E$~($\square$) to the mean-field result $\theta_E =
2z = 2/(2-\gamma)$ (solid line) as a function
of the diffusion exponent~$\gamma$.
} \label{empty_thetaexp} 
\end{figure}

\subsection{Dependence on $\gamma$} \label{depongammasec}

Thus far we have considered only systems with $\gamma = 0$.
All the dynamic scaling properties depend on $\gamma$ and the same is true
for the persistence probabilities as figure~\ref{pers_prob_gamma_fig}
shows. The empty site persistence probability decays
algebraically for all values of $\gamma<2$, above which the system
gels in a finite time.
The analytical prediction, $\theta_E = 2z$, obtained by replacing the
discrete changes in the sizes of clusters besides the persistent one
by a diffusion process, is compared to simulations in
figure~\ref{empty_thetaexp}. The agreement is excellent, which
basically means, that the system self-averages away the
discreteness of the collision times. 
It also tells that there are no nontrivial
correlations such that the clusters surrounding persistent empty
sites would grow differently than the other ones.

The cluster persistence decays algebraically only for $\gamma \ge 0$
and faster than any power of $t$ for $\gamma<0$. The
nonalgebraic behavior 
is not surprising on the basis of the discussion presented in
section~\ref{Clusterpersistencesec}. Although a detailed study of the
cluster persistence confirming the results presented in
section~\ref{Clusterpersistencesec} will be presented
elsewhere~\cite{omapitka}, we briefly comment on the simulational
problems one can encounter.

First, as the persistence exponent is related to the polydispersity
exponent through  $\theta_C = (2-\tau)z$, we 
are faced with same simulation problems than in the
studies of $\tau$ (see e.g.~\cite{Kang:PRA33,Cueille:PRE55}).
Also the crossover time to the asymptotic
behavior is large for $\gamma$ positive but close to zero. In order to
see the real asymptotic behavior there must be very large clusters in
the system.
Consider, for example, $\gamma = 0.2$, which still is not very small. The
diffusion coefficient of a persistent cluster is of order $D_1$ and that
of a typical cluster $D_1S(t)^\gamma$. The ratio of these
two to be of order 10, say, requires $S(t) \approx 10^{1/\gamma} \approx
10^5$, which corresponds to a crossover time $t_{\rm cross} \approx
10^{9}$. Even reaching this in simulations would be hard. This problem
can, of course, be overcome by using a more sophisticated
algorithm. We refer to~\cite{omapitka} for more details on the 
cluster persistence.

\subsection{Intervals between Consecutive Persistent Sites}
\label{sim_intervals} 

In section~\ref{persintandregiondists} we argued that the size
distribution of intervals between persistent quantities would scale
according to $n_X(k;t) = K_X(t)^{-2} f_X(k/K_X(t))$ with
a simple exponential scaling function
[Eq.~\eqref{xXktsimpleexpscal}]. When the corresponding persistence 
probability decays algebraically, $K_X(t) \sim A_X t^{\theta_X}$, with
a nonuniversal amplitude $A_X(\phi)$,  this can be presented also as 
$n_X(k;t) = t^{-2\theta_X} \tilde{f}_X(k/t^{\theta_X})$. 

The difference between these two scaling forms is 
that for the latter the scaling functions 
will not overlap each other for different concentrations due to
the nonuniversal amplitude dependence. Therefore we show for clarity
the scaling plots using the latter formulation. Furthermore, we prefer
to show the scaling of the complement of the cumulative distribution 
\begin{equation}
I_X(k;t) = \sum_{i \ge k} n_X(i;t), \label{compcumdistdef}
\end{equation}
which is much smoother due to the summation. It is
easy to see from the scaling of $n_X(k;t)$ and
Eq.~\eqref{xXktsimpleexpscal} that this 
distribution should scale as $I_X(k;t) = t^{-\theta_X}
g_X(k/t^{\theta_X})$ with $g_X(x) = A_X^{-1} e^{-A_X^{-1}x}$. 

\narrowtext
\begin{figure}
\centering
\includegraphics[width=\linewidth%,draft
]{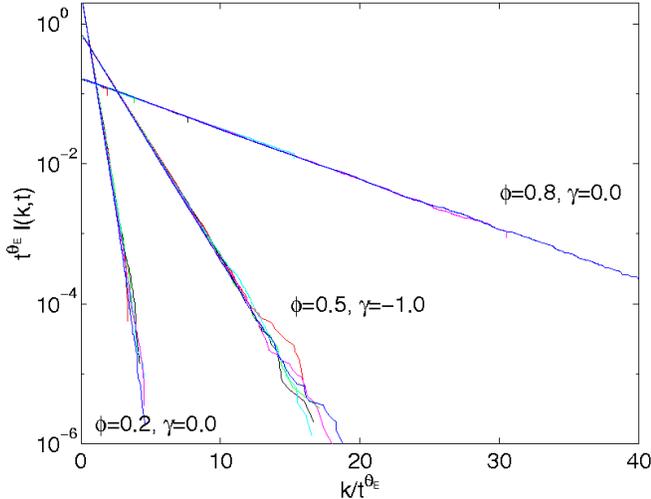}
\caption{Scaling plot for the complement of the cumulative
distribution of intervals between two 
consecutive persistent empty sites at $t=2^9, \ldots, 2^{14}$ for
concentrations and diffusion exponents shown in the figure. The value
used for the persistence exponent $\theta_E=2/(2-\gamma)$. 
} \label{nEkt_scal_plot}
\end{figure}

\narrowtext
\begin{figure}
\centering
\includegraphics[width=\linewidth%,draft
]{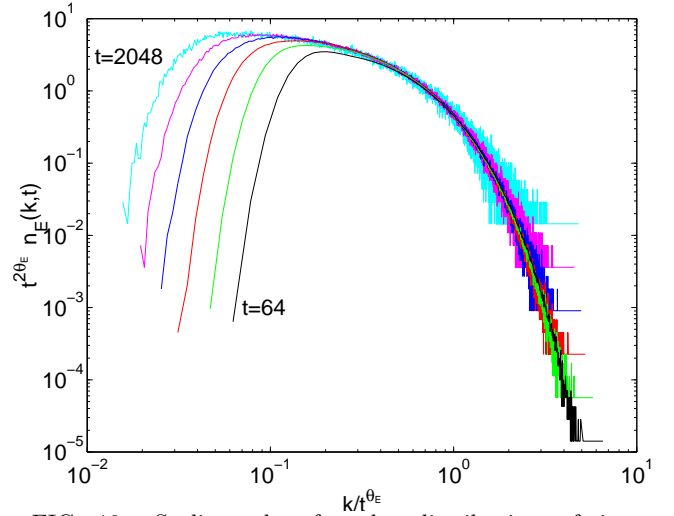}
\caption{Scaling plot for the distribution of intervals between two
consecutive persistent empty sites at $t=2^6, \ldots, 2^{11}$ for
$\phi=0.2$ and $\gamma = 0$.
} \label{nEkt_scal_plot2}
\end{figure}

Figure~\ref{nEkt_scal_plot} shows that the scaling works for 
the empty site persistence and that the scaling function
is an exponential one indeed. The plots for the other persistence
definitions are similar and are not shown. The scaling function is
universal and all the curves in figure~\ref{nEkt_scal_plot} would
overlap each other if one 
would plot $K_E(t) I(k;t)$ as a function of $k/K_E(t)$. Note, that the
diffusion exponent $\gamma$ has no influence on the scaling function. 

Although the
summation in Eq.~\eqref{compcumdistdef} smoothens the data it at the
same time loses information about the small $k/t^{\theta_E}$
behavior. This is  
illustrated in figure~\ref{nEkt_scal_plot2} where no summation is
done. For $k \ll t^{\theta_E}$ the scaling does not work. The
reason is the following. The scaling should work in the limit
$k \to \infty$ and $t \to \infty$ with $y=k/t^{\theta_E}$
fixed. Especially, the condition $k \gg t^z$ should be
satisfied for the two length scales of the problem to be well
separated. Define now  
a time-dependent $y_c(t)$ so that the scaling works for $y >
y_c(t)$. This quantity gets smaller at the same rate as
the curves in figure~\ref{nEkt_scal_plot2} shift towards zero. Our
estimate from the numerical data gives $y_c(t) \sim t^{-0.50 \pm
0.03}$, which is consistent with $y_c(t) \sim t^{-z}$. Thus the
poor scaling in figure~\ref{nEkt_scal_plot2} for $k \ll t^{\theta_E}$
is just a manifestation of the finite time behavior with two competing
length scales, $\LD$ and $\Lp$. This effect vanishes in the scaling
limit. A similar, although not as clear, violation of scaling induced
by the same reason is seen in the study of persistence in the
$q$-state Potts model~\cite{Bray:PRE62}.

\narrowtext
\begin{figure}
\centering
\includegraphics[width=\linewidth%,draft
]{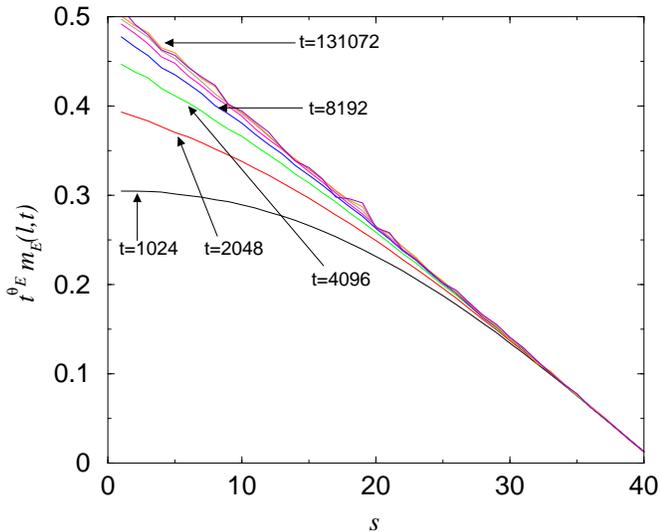}
\caption{The size distributions of persistent
empty regions for $\gamma=0$ and ordered initial conditions with
$s_0=10$ and $\phi = 0.2$. The time instances range from $t=2^{10}$ to $t=2^{17}$.
} \label{region_scaling_2}
\end{figure}

\narrowtext
\begin{figure}
\centering
\includegraphics[width=\linewidth%,draft
]{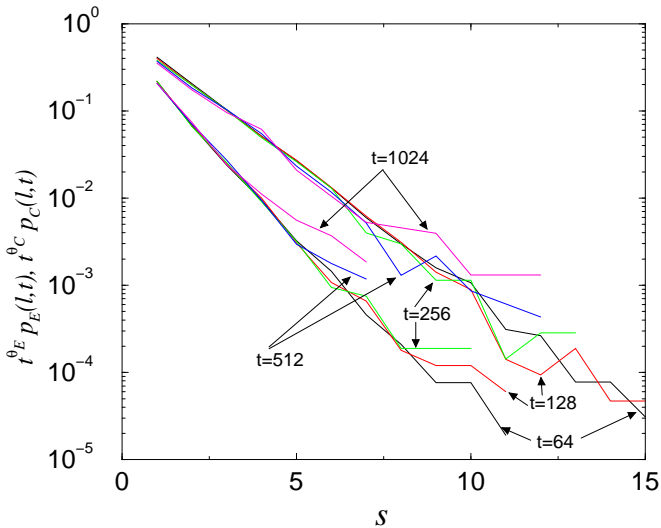}
\caption{The size distributions of persistent
empty regions (upper curves) and clusters (lower curves) for random
initial conditions and
$\gamma=0.75$. The measurement times are shown in the figure. 
} \label{region_scaling_1}
\end{figure}

\subsection{Persistent Regions} \label{sim_regions}

The mean-field arguments used for the empty site exponent predicted
the region size distribution to depend on the initial cluster size
distribution. Calculations indicated
a linear [see Eq.~\eqref{PlTl0eq}] and an exponential
[Eq.~\eqref{Eregiondelta}] decay for the cases
$P(s_0) = \delta(s-s_0)$ and $P(s_0) = s_0^{-1}\exp(-s/s_0)$,
respectively. It was further argued that the dependence on the
diffusion exponent enters only through a multiplicative factor of
$t^{-2/(2-\gamma)}$ in the distributions. This in turn means that the
average region size would approach a constant at late times. 
Figures~\ref{region_scaling_2} and
\ref{region_scaling_1} show the empty region
distributions for these two 
cases and confirm the predictions. 

In Figure~\ref{region_scaling_2}
the smallest times shown are not large enough for the
analysis of section~\ref{emptysiteperssec} to be valid
but the tendency of the distribution to approach a straight line is
clearly visible. The reason why the asymptotic behavior is reached
earlier in Figure~\ref{region_scaling_1} is that the
important measure is not the time itself. Rather it is the ratio
of the diffusion coefficients of clusters, which naturally grows
faster the larger the diffusion exponent $\gamma$ is. 
Note, that in neither of the figures the horizontal axes have 
been scaled, which simply means that the average region size is
constant, \ie, independent of time. This is consistent
with the arguments presented in the
section~\ref{persintandregiondists}. 
The region size distribution between persistent filled sites behaves
in the same way with a concentration dependent exponent~$\theta_F$. 

As discussed at the end of section~\ref{persintandregiondists} the size
distribution of persistent clusters remains unaltered both for
monodisperse initial condition and for $\gamma = 0$. Therefore we only
present the result for random initial conditions and for $\gamma
=0.75$ in figure~\ref{region_scaling_1}. The scaling is the same as it
is for $\gamma = 0$, \ie, the scaling function is a pure exponential and
the average cluster size is asymptotically independent of time.

\section{Conclusions} \label{conclusions}

In this paper we have considered persistence in an aggregation
process, in the simple case of one-dimensional DLCA. The emphasis is
on local properties: cluster, empty site and filled site
persistence together with the corresponding region and interval size
distributions. We have shown that the three persistences are
independent of each other with each behaving in its own, specific way. 

The perhaps most natural choice, the probability that a site has remained
in the same state (either filled or empty), turns out to be nonuniversal.
Nevertheless, only one of the two contributing probabilities, the filled
site persistence, is responsible for this fact. The other one, the
empty site persistence, is universal. In the DCLA
the difference in the dynamics of empty and filled sites implies
that there is no symmetry that would produce similarity in the
persistence behaviors. The reason why the cluster persistence is a
third independent quantity derives from the fact that it is a
cluster property whereas the two other persistence definitions are
considered at a fixed point in space. 

To summarize,
the filled site persistence decays asymptotically algebraically for
$\gamma < 2$ and for all concentrations: $P_F(t) \sim 
t^{-\theta_F}$. The persistence exponent $\theta_F$ depends on
concentration and is 
therefore nonuniversal. At low concentrations the filled site
persistence decays as a stretched exponential upto a concentration
dependent crossover time after which the collisions between clusters
become important and start to dominate the persistence behavior. 

The empty site persistence is universal. Although we have not been
able to prove this, the universality is supported
both by mean-field type continuum arguments and by computer
simulations. The former leads to a relatively simple relation between
the persistence exponent and the dynamic exponent, $\theta_E =
2z$, verified by simulations. This is one of the few
examples~\cite{Bray:PRE62}, where the inequality $\theta > 
zd$, where $d$ is the spatial dimension, is fulfilled. The consequence
of this is that the persistent empty regions do not have a fractal
character. This is not true for example for the persistent regions in the
Ising~\cite{Jain:JPHYSA33,Bray:PRE62} or
diffusion--annihilation~\cite{Manoj:JPA33L} models. The fact that
$\theta_E$ is notably larger than $z$, makes the separation of
the diffusive and persistence length scales clearly visible
in the DLCA.

The cluster persistence probability decays algebraically only for $0
\le \gamma <2$ and as stretched exponentially for $\gamma <
0$. For $0 < \gamma <2$ the persistence exponent is given by
$\theta_C(\gamma) = 2z = 
2/(2-\gamma)$ and it is discontinuous as $\gamma \to 0^+$ since
$\theta_C(0) = \frac{3}{2}$. All these results are in close connection
with the scaling of the cluster size distribution, especially with the
small $x$ decay of the scaling function $f(x)$. 
In fact, there is a scaling relation between the exponents
$\theta_C = (2-\tau)z$. This 
independence of the cluster persistence probability on concentration
and the strong support of the universal dynamic scaling 
behavior of the cluster size distribution~\cite{Hellen:PRE62} speak
for the universality of the cluster persistence. The scaling relation
together with the result for $\theta_C$ offers a new way to approach
the determination of $\tau$ for DLCA.

It is worth emphasizing that the universal empty and cluster
persistences decay with a same exponent $\theta_E = \theta_C = 2z =
2/(2-\gamma)$ for $0 < \gamma < 2$ but have nothing to do with each
other for $\gamma < 0 $. In fact, for $\gamma \ge 0$ we can write
$P_C(t) \sim [P_E(t)]^\Gamma$, where $\Gamma$ takes the values $3/2$
and $1$ for $\gamma = 0$ and $0< \gamma < 2$, respectively. A similar
relation, $P_C(t) \sim [P_E(t)]^\Gamma$, with a non-trivial $\Gamma$
has been observed also for noninteracting random
walkers~\cite{ODonoghue:CM1} and for the Potts
model~\cite{ODonoghue:CM2} in one dimension. In DLCA the same
persistence exponents, \ie, $\Gamma =1$, for positive $\gamma$ are 
due to the fact that the clusters giving the dominant
contribution to the cluster persistence are those which remain
stationary. For other values of $\Gamma$ the interpretation is more
opaque. 

Although the persistences decay in a different manner (algebraic or
faster) the intervals between persistent regions scale similarly for
all cases: $n_X(k;t) = K_X(t)^{-2} f_X(t/K_X(t))$. This is due
to the relation $\theta_X > z$, which causes 
the persistent regions to be well separated at late times. Therefore
the regions survive independently and the intervals size distribution
is a simple exponential $f_X(y) = e^{-y}$.

The size distributions of persistent regions depend on the initial
cluster size distribution. They do not depend on the diffusion
exponent which only enters through a multiplicative time dependence
on the distribution, for example, $p_E(l,t) \sim t^{-2/(2-\gamma)}
\tilde{p}_E(l)$. This leads to a constant average region size at late
times, which can be estimated from the knowledge of the initial
cluster size distribution. 

In conclusion, we have presented a rather comprehensive
study of various local persistence probabilities in the
one-dimensional DLCA. Our study is of interest also for the sake of practical
realizations. Aggregation processes are plentiful, and
all of the definitions - whether a point in space is 
occupied or not by a cluster, or whether clusters
survive intact - might well be possible to measure in
experimentally. It is interesting also to note that 
when the decay of a persistence probability turns out
to be algebraic and universal, the exponent is always directly in some
relation to the dynamical exponent $z$ of the aggregation
process. It is an obvious question to ask how the various quantities
work out in higher dimensions. This generalization seems both challenging
and interesting.

\section{Acknowledgments}

The authors thank P.~E.~Salmi for numerous discussions. 
E.~K.~O.~H. also thanks 
%A.~J.~Bray and P.~L.~Krapivsky for helpful discussions and 
S. Majaniemi for helpful remarks.
This research has been supported by the Academy of Finland's Center of
Excellence program.

%\bibliographystyle{prsty}
%\bibliography{BIBTEX/omatabbrev,BIBTEX/pers}

\end{multicols}

\end{document}